\documentclass[10pt,preprint2]{aastex}
\def\figdir{.}

\def\ion#1#2{\mbox{\rm #1\sc #2}}

\def\HII{{\ion{H}{ii}}}

\def\dim#1{\mbox{\,#1}}
\def\lya{Lyman-$\alpha$}
\def\lyb{Lyman-$\beta$}
\def\lyg{Lyman-$\gamma$}
\def\we{we}
\def\us{us}
\def\our{our}
\def\wmap{{\it WMAP\/}}
\def\tgp{\tau_{\rm GP}}
\def\tt{\tau_{\rm T}}

\def\figname#1{\figdir/#1}

\begin{document}

\title{Cosmic Reionization Redux}
\author{Nickolay Y.\ Gnedin\altaffilmark{1,2} and Xiaohui Fan\altaffilmark{3}}
\altaffiltext{1}{Particle Astrophysics Center,
Fermi National Accelerator Laboratory, Batavia, IL 60510, USA; gnedin@fnal.gov}
\altaffiltext{2}{Department of Astronomy \& Astrophysics, The
  University of Chicago, Chicago, IL 60637 USA} 
\altaffiltext{3}{Steward Observatory, The University of Arizona, Tucson, AZ 85721; fan@as.arizona.edu}

\def\dim#1{\mbox{\,#1}}

\label{firstpage}

\begin{abstract}
We show that numerical simulations of reionization that resolve
the Lyman Limit systems (and, thus, correctly count absorptions of
ionizing photons) have converged to about 10\% level for $5<z<6.2$ and
are in reasonable agreement (within 10\%) with the SDSS data in this redshift
interval. The SDSS data thus constraint the redshift of overlap of
cosmic \HII\ regions to $z_{\rm OVL} = 6.1\pm0.15$. At higher
redshifts, the simulations are far from convergence on the mean
Gunn-Peterson optical depth, but achieve good convergence for the mean
neutral hydrogen fraction. The simulations that fit the SDSS data,
however, do not have 
nearly enough resolution to resolve the earliest episodes of star
formation, and are very far from converging on the precise value of
the optical depth to Thompson scattering - any value between 6 and
10\% is possible, depending on the convergence rate of the simulations
and the fractional contribution of PopIII stars. This is generally
consistent with the third-year \wmap\ results, but much higher
resolution simulation are required to come up with the sufficiently
precise value for the Thompson optical depth that can be statistically
compared with the \wmap\ data.
\end{abstract}


\section{Introduction}

Bad things comes in threes - but good things sometimes do too. At
least, in twos. In the reionization research, the two are the latest
analysis of the \lya\ absorption in the spectra of 19 highest redshift
SDSS quasars by \citet{fea06a} and the large downward revision of the
\wmap-measured value for the Gunn-Peterson optical depth
\citep{pea06a,sea06a}. 

The latest SDSS data are consistent with the end of reionization epoch
at $z\sim6$ \citep{fea06a}. The downward revision of the \wmap\
measurement now eliminates any need for complex scenarios of early
reionization \citep[c.f.][and references therein]{mcsf05a}, which had
to invoke unknown or weakly constrained physics and which were at odds
with the simplest reionization scenario in which normal PopII stars
are the dominant source of ionizing photons between $z\sim20$ and
$z\sim5$. 

In this simplest scenario \citep{g00a,g04a,rf06a} - which we call the ``Minimal
Reionization Model'' - PopIII stars and other exotic objects play an
important but not a dominant role in reionizing the universe by
$z\sim6$. The bulk of cosmic ionizations are produced by normal PopII
stars in sufficiently massive galaxies (like the ones observed in UDF
and GOODS). As these galaxies form, they create ionized (\HII) regions
around them, which continue to expand. The universe is half-ionized
(by volume) at $z\sim8-9$, and shortly before $z=6$ most of cosmic
\HII\ regions merge during a relatively short period called
``overlap'', completing the process of reionization of the universe. 

Since it now appears that the Minimal Reionization Model provides
the best theoretical framework for the existing data, it would
make sense to refine its predictions and expectations to make a more
precise comparison with the SDSS and \wmap\ data. It is particularly
timely now, since several different groups make simulations of
reionization, and these simulations rarely agree with each other.

In this paper we revisit the agreement with the data (or lack of it) for the
simulations reported in \citet{g04a}, as well as for the latest
simulation that is substantially larger and has much higher spatial
resolution. The existence of a set of simulations with varied box
sizes and mass and spatial resolutions
allows \us\ not only to compare the data to the model, but to also estimate the
level of numerical convergence of the simulations, and, thus, make a
rough estimate of the current theoretical uncertainty.

\section{Simulations}

In this paper \we\ use three simulations that have different spatial
and mass resolutions and different sizes of the simulation volumes,
which allows \us\ to estimate the degree of numerical convergence for
different physical quantities. All simulations have been performed
with the Softened Lagrangian Hydrodynamics (SLH) code \citep{g00a}
using the Optically Thin Eddington Tensor (OTVET) approximation
\citep{ga01a} for following the time-dependent and spatial-variable
transfer of ionizing radiation in 3D. All three simulations use the
same cosmological parameters as reported in
\citet{svp03a}\footnote{$\Omega_M=0.27$, $\Omega_\Lambda=0.73$,
  $h=0.71$, $\Omega_B=0.04$, $n_S=1$.}

\begin{table}[t]
\caption{Simulation Parameters\label{sim}}
\begin{tabular}{lccccc}
\tableline
Run &
$L$\tablenotemark{a} &
$\Delta x$\tablenotemark{b} &
$\Delta M$\tablenotemark{c} &
$z_{\rm OVL}\tablenotemark{d}$ &
$\tt$ \\
\tableline
\tableline
L4N128 & 4 & 1    & $3.2\times10^6$ & 6.2 & 0.051 \\
L8N128 & 8 & 2    & $2.6\times10^7$ & 6.2 & 0.048 \\
L8N256 & 8 & 0.64 & $3.2\times10^6$ & 6.2 & 0.056 \\
\tableline
\end{tabular}
\tablenotetext{a}{Size of the computational box in $h^{-1}\dim{Mpc}$.}
\tablenotetext{b}{Spatial resolution in comoving $h^{-1}\dim{kpc}$.}
\tablenotetext{c}{Mass resolution in $M_\odot$.}
\tablenotetext{d}{By construction.}
\end{table}
The numerical parameters of the simulations are listed in Table
\ref{sim}. The first two simulations are ones labeled ``A4'' and ``A8'' in
\citet{g04a} and \citet{fea06a}, while the third one is a new
simulation with $256^3$ dark matter particles, the same number of
baryonic quasi-Lagrangian cells, and about 1{,}300{,}000 stellar
particles that have been forming continuously during the
simulation. In order to make references to specific simulations
transparent, \we\ label each simulation with a letter L followed by
the value of the linear size of the computational volume (measured in
$h^{-1}\dim{Mpc}$ in comoving reference frame), followed by the letter
N and the number of dark matter particles (or baryonic cells) along
one direction (128 or 256). For example, L4N128 means a simulation
with $4h^{-1}\dim{Mpc}$ box size and with $128^3$ dark matter
particles and an equal number of baryonic cells. 

Table \ref{sim} also gives the nominal spatial resolution of the
simulations (as measured by the value of Plummer softening length -
the real resolution being a factor of 2-4 worse). Because of the
specifics of the SLH method, a mesh with a larger number of cells can
be deformed more, so the spatial resolution of the large L8N256
simulation is, in fact, higher than the spatial resolution of the
L4N128 simulation, even if the mass resolutions of both simulations are
identical. 

Finally, \we\ also list in Table \ref{sim} values for the redshift of overlap,
$z_{\rm OVL}$ \citep{g04a}, and the Thompson optical depth, $\tt$, that
we use below.

\begin{figure}[t]
\plotone{\figname{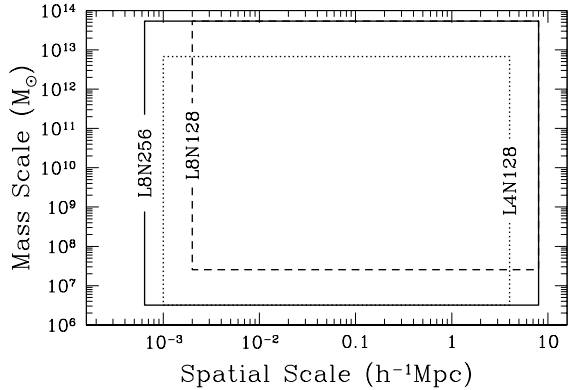}}
\caption{A graphical representation of the spatial and mass
  resolutions for the three simulations we use in this paper. The
  horizontal axis is the spatial scale and the vertical axis is the
  mass scale. Spatial and mass scales resolved in the three
  simulations are shown as three rectangles.}
\label{figSS}
\end{figure}
It is convenient to represent the numerical resolution of a given
simulation as a rectangle in the spatial scale vs mass scale plane, as
shown in Figure \ref{figSS}. The horizontal axis is the spatial scale
({\it not\/} a given spatial direction), and since a simulation has
a finite spatial resolution and a finite box size, it is limited
along the spatial axis at both ends. The same is true for the mass
scale, so in the spatial scale - mass scale plane the simulation is
represented by a rectangle. The three simulations are shown in Fig.\
\ref{figSS}. The L8N256 simulation has the largest spatial and mass
dynamic range. The L8N128 run has the same box size as the L8N256, but
lower spatial and mass resolution, while the L4N128 run has the same
mass resolution as the L8N256 run, but lower box size (and somewhat
lower spatial resolution, as explained above).

Because of the computational expense, and because \our\ simulations
become unreliable for $z<5$ (as we explain below), the L8N256 run has been
continued only until $z\sim5$.

All simulations have been adjusted to best fit the mean transmitted
flux data as explained in \citet{g04a}.

\section{How to Simulate Reionization Correctly}

Before \we\ can compare the simulations and the data, it is important
to explain why \our\ simulations are adequate for modeling
reionization, despite the limited box size. Modeling reionization,
after all, is all about counting absorptions of ionizing photons
correctly. It is well known that after reionization, absorption of
ionizing photons is dominated by the Lyman Limit systems
\citep{m03a}. Obviously, the same should be true inside large enough
\HII\ regions even during reionization, so resolving the Lyman Limit
systems is crucial for counting the absorptions of ionized photons
correctly during and, perhaps, even before the overlap stage of reionization.

\begin{figure}[t]
\plotone{\figname{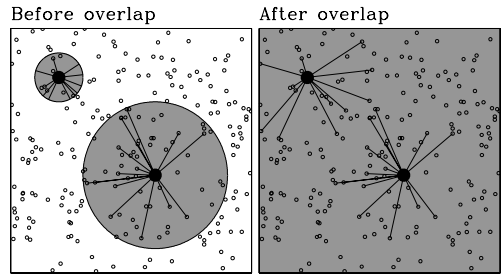}}
\caption{A sketch of two ionizing sources before and after
  overlap. Before overlap the mean free path for photons emitted by a
  weak source (in the upper left corner) is limited by the size of its
  \HII\ region (gray area), while the strong source (in the lower
  right corner) has reached its Stromgen sphere and all photons it
  emits are absorbed in Lyman Limit systems (small open cicles). After
  overlap, Lyman Limit system determine the mean free path for all
  sources.}
\label{figCT}
\end{figure}
To illustrate this point, we show in Figure \ref{figCT} a sketch of
\HII\ regions around two sources (a weak one and a strong one) before and
after the overlap. Before overlap, most of photons emitted by a sufficiently
weak source are absorbed by the neutral IGM just outside its I-front,
while a sufficiently strong source, whose I-front reached the size
comparable to the mean free path of ionizing photons inside it,
effectively reaches its ``Stromgen sphere'', because inside its \HII\
region ionizations balance recombinations\footnote{Note, that \we\ use
  the term ``Stromgen sphere'' here differently from \citet{sg87a}, who
  did not account for Lyman Limit systems and thus concluded that
  cosmic \HII\ regions never reach their Stromgen spheres.}.
After the overlap, all sources reach their ``Stromgen spheres'', with
ionization nearly balancing recombinations within the Lyman Limit
systems \citep{mhr00a}.

The exact nature of Lyman Limit system is still rather poorly
understood, but Lyman Limit systems in a simulation can still be
studied by closely mimicking the observational process. The simulations
\we\ use in this paper do indeed resolve the Lyman Limit system
\citep{kg06b} - in fact, they do it only too well by overpredicting the
observed numbers of Lyman Limit systems at $z\sim4$ by about a factor
of 2 and by smaller factors at higher redshifts. The discrepancy is
because the simulations do not properly include the ionizing radiation from
quasars, and they become inadequate for describing the
ionization state of the IGM for $z\la5$. This inadequacy at lower
redshifts also becomes apparent in \our\ results presented below.

Based on the analysis of the Lyman Limit systems from cosmological
simulations \we\ use here \citep{kg06b} it appears that spatial
resolution of at least 1-2 proper kpc is required to model the
Lyman Limit systems correctly. Simulations of poorer spatial
resolution would not be adequate for modeling cosmic reionization
because they would underestimate absorption of ionizing radiation by
a large factor.

\section{Results}

\begin{figure}[t]
\plotone{\figname{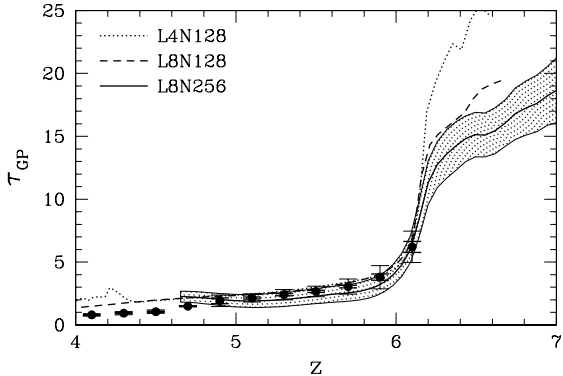}}
\caption{The mean Gunn-Peterson optical depth as a function of redshift for
  the simulations (gray lines) and data (black symbols). The dotted
  line shows the L4N128 run, the dashed line is for the L8N128 run,
  and the solid line is for our largest L8N256 run. In the latter
  case, the rms variation around the mean is shown as a hatched
  band. Filled circles show the observational data from
  \protect{\citet{fea06a}} with both rms dispersion (thin error-bars)
  and errors of the mean (thick error-bars).} 
\label{figOD}
\end{figure}
We first show in Figure \ref{figOD} the evolution of the mean
Gunn-Peterson optical depth, defined as
\begin{equation}
  \tgp = -\log\left(\langle F(z)\rangle\right),
\end{equation}
where $\langle F(z)\rangle$ is the mean (i.e.\ averaged over all
possible directions) transmitted flux in the hydrogen \lya\ transition
at a given redshift. It is important to underscore that the mean
Gunn-Peterson optical depth is {\it not\/} an average of anything, and,
following \citet{fea06a}, we emphasize it by {\it not\/} using the
overbar or the averaging operator $\langle\rangle$ in its definition.

The simulations are within about 10\% of the data and within 10\% of
each other in the redshift interval
$5\la z\la6.2$, but the agreement with the data 
for the $5.8< z<6.2$ interval is by
construction - \we\ adjust the effective emissivity parameter in the
simulations to fit the observed data, as is explained in
\citet{g04a}. The agreement in the interval $5<z<5.8$ is real - it
reflects the fact that the ionization state of cosmic gas in the 
simulations is broadly consistent with the observational data (the
following figures provide more detail on the level of agreement
between the simulations and the data). 

The fact that simulations with various box sizes and resolutions
differ from each by about 10\% demonstrates that our numerical results
have not converged to that level of precision.

Formally, none of \our\ three simulations is actually
consistent with the data at $z<5.8$ in the statistical sense - the simulations
lie within the rms fluctuation in the Gunn-Peterson optical depth for
a range of redshifts, but outside the formal errors in the mean value
of $\tgp$, which are about 5\% for $z<5.8$. However, the systematic
errors in the measurement (continuum fitting and possible
contaminations from BAL/metal absorption in \lya\ forest region) are
likely to be of the order of 5 to 10\%, so the agreement between the
simulations and the data at 10\% level is satisfactory at present and
is an encouraging confirmation of the Minimal Reionization
Model. 

We also emphasize the challenge facing the simulations in the years to
come. As the random and especially systematic errors in the data are
reduced, the current simulations will be hard pressed to fit the data
at, say, 3 to 5\% level. The future simulations will have to follow the
radiative transfer correctly in fine detail to fit the observations,
and this sensitivity to details offers a tremendous opportunities to
learn about nature and distribution of ionizing sources at $z<5.8$.

As \we\ have mentioned above, the simulations become inadequate for $z<5$
because they do not properly include the ionizing radiation from
quasars. As the result, there are more Lyman Limit systems in the
simulation \citep{kg06b}, the IGM in the
simulations is more neutral, and the mean free path of ionizing
radiation for $z<5$ is shorter than are actually observed.

At $z>6.2$ the simulations show a marked lack of convergence in the
mean Gunn-Peterson optical depth. This
difference cannot be explained away entirely by cosmic variance, since
the rms fluctuations in, say, the L8N256 run are smaller than the
difference between various runs. 

This difference, however, is not surprising. After all, a mean
Gunn-Peterson optical depth of 10 corresponds to the mean transmitted
flux of only $10^{-5}$ - in a simulation with $128^3\sim2$ million
cells only 20 transparent ($F=1$) cells in the complete opaque ($F=0$)
medium would produce such a small mean transmitted flux. Thus, even
discreetness effects in the simulation become important at these low
values of the mean transmitted flux, in addition to usual effects of
limited resolution, poorly known physics, lack of numerical
convergence, etc - this is merely a statement of the fact that before
the overlap any measurable transmission comes from the very tail of
density and neutral fraction fluctuations, and, thus, is exceedingly
difficult to simulate correctly. The latter statement only applies to
the transmitted flux in the hydrogen \lya\ transition, other
quantities like mean fractions or photoionization rates can be modeled
more reliably before the overlap.

\begin{figure}[t]
\plotone{\figname{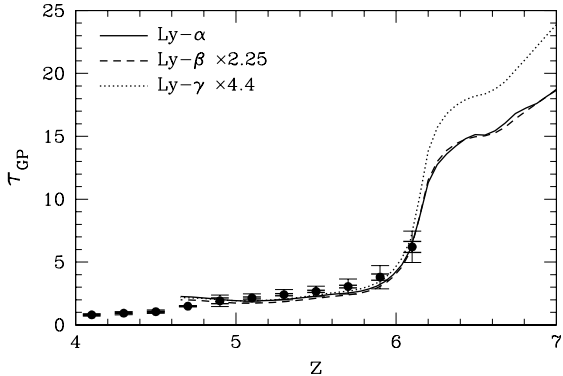}}
\caption{The mean Gunn-Peterson optical depth as a function of redshift for
  the L8N256 simulation (gray lines) and data (black symbols). The solid
  line shows the \lya\ measurement (the same line as in Fig.\
  \protect{\ref{figOD}}), and dashed and dotted lines show the \lyb\
  and \lyg\ measurements respectively, rescaled to \lya\ by factors
  2.25 and 4.4, as used by \protect{\citet{fea06a}}.}
\label{figBG}
\end{figure}
It is also important to emphasize that the direct comparison between
the simulations and the data for $\tgp>10$ is highly
non-trivial. The observational data constrain the \lya\ optical depth
directly only for $\tgp\la7$, and, for higher values, properly
rescaled constraints from \lyb\ and \lyg\ transitions are used. When
\we\ repeat the \citet{fea06a} rescaling procedure with \our\ L8N256
simulation, \we\ get a good agreement between the \lyb\ and \lya\ for
the scaling factor of 2.25, but get a higher optical depth for the
\lyg\ transition if we adopt \citet{fea06a} scaling value of 4.4,
while a smaller value of about 3.5 gives the best agreement between the
\lyg\ and \lya\ measurements.

\begin{figure}[t]
\plotone{\figname{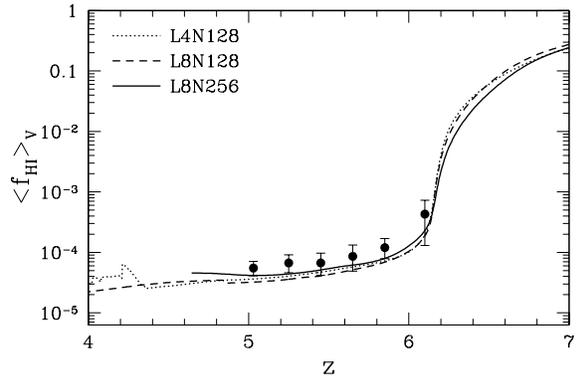}}
\caption{The mean volume-weighted neutral hydrogen fraction as a
  function of redshift for the simulations and data. The line and
  symbol markings are as in Fig.\ \protect{\ref{figOD}}.}
\label{figXV}
\end{figure}
Figure \ref{figXV} shows the comparison between the SDSS data and the
simulations for the mean volume weighted neutral fraction\footnote{The
  comparison is much more difficult for the mean mass weighted neutral
  fraction, because the simulations do not resolve Damped \lya\
  systems, that are known to contain most of neutral gas in the
  universe at $z\la4$, and are likely to contain at least a
  substantial fraction of all neutral gas at $z\la6$ as well.}
as a function of redshift. In this quantity, the convergence of the
simulations is much better, and the data are consistent (at 10 to 20\%
level) with the simulation results. This agreement is not surprising,
given the agreement in the mean Gunn-Peterson optical depth - the
neutral fraction (and the mean free path that we discuss below) is not
measured directly from the observations, but is rather derived
from the Gunn-Petrson optical depth measurement based on assumed density
distribution of the IGM and photoionization equilibrium.

\begin{figure}[t]
\plotone{\figname{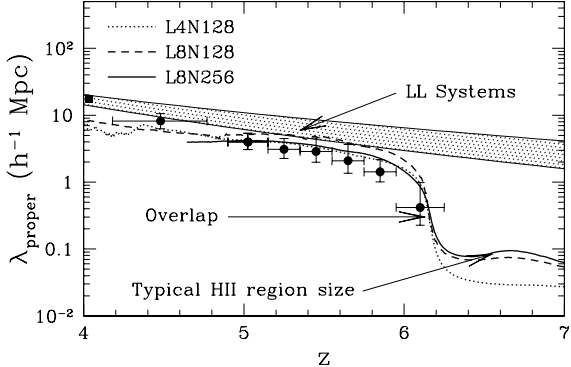}}
\caption{The mean free path of ionizing radiation as a
  function of redshift for the simulations and data. The line and
  symbol markings are as in Fig.\ \protect{\ref{figOD}}. The filled
  black square at $z=4$ shows the observational determination of the
  mean free path by \protect{\citet{m03a}}, and the hatched black band
  is the extrapolation of the observed evolution of the Lyman Limit
  systems from \citet{smih94a} to $z>4$.}
\label{figMP}
\end{figure}
In Figure \ref{figMP} \we\ show the evolution of the mean free path of
the Lyman Limit photons in the simulations and in the SDSS data, as
well as a plausible extrapolation to high redshift of the mean free
path from the Lyman Limit systems alone \citep{smih94a,m03a}. As we
have mentioned before, the simulations slightly underestimate the mean
free path because they overpredict the abundance of the Lyman Limit
systems up to a factor of 2 \citep{kg06b}. The SDSS data only go up to
the overlap epoch at $z\approx6.2$, and are not yet able to probe the
approach to the overlap at $z>6.2$.

The SDSS estimate of the mean free path lies somewhat below the
extrapolation from the Lyman Limit system measurement at $z=4$. This
difference is most likely not significant, as the extrapolation
assumes a power-law evolution of Lyman Limit systems in redshift,
which is too simple to be precise. But the general agreement between the
extrapolation and the SDSS result indicates that the mean free
path of Lyman Limit photons is limited by the Lyman Limit systems for
$z<5.8$, as is generally believed.

In the simulations, the strong deviation from the power-law evolution
of the mean transmitted path with redshift at $z>6$ indicates the
overlap stage of reionization - during which the mean free path is
still determined by a typical size of \HII\ regions rather than by the
Lyman Limit systems. If we take the agreement between the SDSS data
and the simulations for $z<6.2$ as an indication that the rapid
decrease in the SDSS estimate of the mean transmitted path indeed
corresponds to the overlap of cosmic \HII\ regions, then we can adopt
the last SDSS data point as a plausible estimate of the time of
overlap (formally defined as the moment when the mean free path grows
most rapidly), giving $z_{\rm OVL}=6.1\pm0.15$.

In particular, the SDSS data limit the typical size of a
cosmic \HII\ region to less than about $1h^{-1}$ proper Mpc at
$z\approx 6.2$.

\begin{figure}[t]
\plotone{\figname{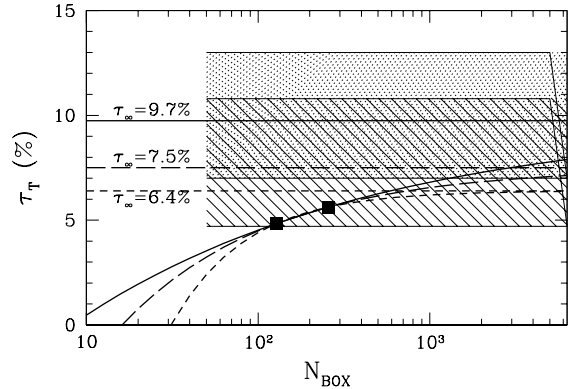}}
\caption{The Thompson optical depth as a function of resolution for the
  L8N128 and L8N256 simulations (black squares). The gray and black
  hatched bands show the constraints on $\tt$ from the \wmap\
  polarization measurement alone ($\tt=0.10\pm0.03$,
  \protect{\citet{pea06a}}) and from the combined \wmap+SDSS data
  ($\tt=0.08\pm0.03$, \protect{\citet{sea06a}}) respectively. Black
  dotted, dashed, and solid curved 
  lines show the fits to the simulation results in the form
  $\tt(N)=\tau_\infty+A/N^\alpha$ for $\alpha=1$, $1/2$, and $1/4$
  respectively, while horizontal lines of the same type give the
  converged values $\tau_\infty$.}
\label{figTT}
\end{figure}
Finally, we can compare the optical depth to Thompson scattering,
$\tt$, to the \wmap\ measurements. These values for the three
simulations are given in Table \ref{sim} and also shown in Figure
\ref{figTT}. As one can see, these values are far from being
converged, because the smallest objects that form first (and are
missed in finite resolution simulations) can contribute significantly
to the Thompson optical depth. In numerical analysis it is customary
to estimate the converged result for any physical quantity $C$ from a
set of simulations with varied mesh size $N$ by fitting the computed
values of $C(N)$ with a functional form
\begin{equation}
  C(N) = C_\infty + \frac{A}{N^\alpha}.
  \label{converge}
\end{equation}
In order to determine three coefficients $C_\infty$, $A$, and
$\alpha$, three simulations with three different mesh sizes are
required. Unfortunately, in \our\ case it is not possible to get
another simulation with a significantly different value of $N$: running
a $512^3$ SLH simulation is completely unfeasible now, and a $64^3$
simulation is so small that it does not fit the SDSS data at
all. Also, using a value of $N$ somewhere between $128$ and $256$
would not help, as Fig.\ \ref{figTT} illustrates. 

In order to illustrate the inadequacy of \our\ simulations for giving
an accurate estimate of $\tt$, \we\ fitted equation (\ref{converge})
for the L8N128 and L8N256 simulations (since the resolution is the most
important quantity for modeling the highest redshift star formation)
keeping $\alpha$ as a free parameter, and fits for $\alpha=1$, $1/2$,
and $1/4$ are shown in Fig.\ \ref{figTT}. As anyone can see, the
simulations are non-conclusive: depending on the rate of numerical
convergence, the simulations can produce any value for $\tt$ between
about 6 and 10\%.

In addition, a significant (but not dominant) contribution from PopIII
stars is also possible.

\section{Conclusions}

We have shown that numerical simulations of reionization that resolve
the Lyman Limit systems (and, thus, correctly count absorptions of
ionizing photons) are close (within 10\% or so) to the SDSS data for a
variety of measured and deduced quantities for $z<6.2$. The SDSS data
thus constraint the redshift of overlap to $z_{\rm OVL} = 6.1\pm0.15$.

The mean free path of ionizing photons is limited by the Lyman Limit
systems at $z<6.0$, and is smaller at higher redshifts, reflecting a
finite size of a typical cosmic \HII\ region (less than $1h^{-1}$
proper Mpc at $z=6.2$).

Our simulations, however, do not have
nearly enough resolution to resolve the earliest episodes of star
formation, and are very far from converging on the precise value of
the optical depth to Thompson scattering - any value between 6 and
10\% is possible, depending on the convergence rate of the simulations
and the fractional contribution of PopIII stars. This is generally
consistent with the third-year \wmap\ results, but much higher
resolution simulation are required to come up with the sufficiently
precise value for the Thompson optical depth that can be statistically
compared with the \wmap\ data.

While our simulations agree with the SDSS data within about 10\%
for $5<z<6.2$, the level of numerical convergence of simulations in
this redshift interval is 10\% at best. As the data improve (mostly
by reducing the systematic errors), the constraining power of the
SDSS data will be sufficient to place non-trivial demands on the
degree of realism of any simulation that attempts to fit the data
statistically. Not only the cosmological parameters in the simulation
must be precise enough, but it is likely that fine details of
radiative transfer (the relative roles of quasars and galaxies, their
spatial clustering, accuracy of numerical schemes, etc) have to be
done accurately as well. These requirements will present a challenge
and a motivation for the future theoretical work on modeling cosmic
reionization with greater precision than has been possible so far.

\bibliography{rei5}

\end{document}